\documentclass[aps,prl,reprint,amsmath,amssymb]{revtex4-2}
\usepackage[bookmarks=false,linkcolor=NavyBlue,urlcolor=NavyBlue,colorlinks,citecolor=NavyBlue]{hyperref}
\usepackage{graphicx}
\usepackage{amsmath}
\usepackage[svgnames]{xcolor}
\usepackage{enumerate}
\usepackage[normalem]{ulem}
\newcommand{\figref}[2]{\hyperref[#1]{\ref{#1}(#2)}}
\newcommand{\figsref}[2]{\hyperref[#1]{\ref{#1}#2}}
\newcommand{\Weizmann}{Department of Condensed Matter Physics, Weizmann Institute of Science, Rehovot, Israel 7610001}

\begin{document}

\title{One-dimensional topological superconductivity based entirely on phase control}

\author{Omri Lesser}
\affiliation{\Weizmann}

\author{Yuval Oreg}
\affiliation{\Weizmann}

\author{Ady Stern}
\affiliation{\Weizmann}

\begin{abstract}
Topological superconductivity in one dimension requires time-reversal symmetry breaking, but at the same time it is hindered by external magnetic fields.
We offer a general prescription for inducing topological superconductivity in planar superconductor--normal--superconductor--normal--superconductor (SNSNS) Josephson junctions without applying any magnetic fields on the junctions.
Our platform relies on two key ingredients: the three parallel superconductors form two SNS junctions with phase winding, and the Fermi velocities for the two spin branches transverse to the junction must be different from one another.
The two phase differences between the three superconductors define a parameter plane which includes large topological regions.
We analytically derive the critical curves where the topological phase transitions occur, and corroborate the result with a numerical calculation based on a tight-binding model.
We further propose material platforms with unequal Fermi velocities, establishing the experimental feasibility of our approach.
\end{abstract}
\maketitle

\emph{Introduction.---}Topological superconductivity is a novel phase of matter with fascinating edge physics~\cite{qi_topological_2011,alicea_new_2012,leijnse_introduction_2012,bernevig_topological_2013}.
In one dimension, topological superconductors host Majorana zero modes at their ends, which possess exotic exchange properties~\cite{kitaev_unpaired_2001}.
In experimental setups, attempts to engineer topological superconductors rely on proximity-coupling to non-topological $s$-wave superconductors and employing strong spin-orbit coupling (SOC) to separate the two spin species.
Furthermore, time-reversal symmetry has to be broken to lift the Kramers degeneracy.
When judiciously combined, these three ingredients --- conventional superconductivity, a spin-rotation mechanism, and time-reversal symmetry breaking --- make the low-lying energy band effectively spinless while maintaining superconducting pairing, thus giving rise to spinless topological superconductivity.

Much effort has been devoted to inducing topological superconductivity in various experimental platforms~\cite{lutchyn_majorana_2018,flensberg_engineered_2021}.
One of the first proposals~\cite{fu_superconducting_2008} utilized the surface of topological insulators~\cite{hasan_colloquium_2010,asboth_short_2016} in proximity to a superconductor.
Other prominent proposed Majorana platforms include semiconductor--ferromagnet heterostructures~\cite{sau_generic_2010}, quantum wells with an in-plane magnetic field~\cite{alicea_majorana_2010}, and chains of magnetic adatoms on superconductors~\cite{pientka_topological_2013,nadj-perge_observation_2014}.
Early on, semiconductor--superconductor nanowires were put forward as an accessible platform~\cite{lutchyn_majorana_2010,oreg_helical_2010}.
Nanowires have since then been vigorously studied: theoretical extensions of the original models were made to include current biasing~\cite{romito_manipulating_2012}, disorder~\cite{pan_physical_2020,pan_disorder_2021}, electrostatics~\cite{escribano_effects_2019}, full-shell nanowires~\cite{stanescu_robust_2018,vaitiekenas_flux-induced_2020}, as well as electron--electron interactions~\cite{sela_majorana_2011,stoudenmire_interaction_2011,oreg_fractional_2014}.
On the experimental side, several groups have reported possible signatures of Majorana zero modes in proximitized nanowires~\cite{mourik_signatures_2012,das_zero-bias_2012,albrecht_exponential_2016,deng_majorana_2016,grivnin_concomitant_2019,vaitiekenas_flux-induced_2020,vaitiekenas_zero-bias_2020}, but they are not entirely definitive~\cite{vuik_reproducing_2019,pan_physical_2020,hess_local_2021}.

The experimental drawbacks of the nanowire platform is the need for large magnetic fields,
which hinder superconductivity and create unwanted sub-gap states~\cite{sabonis_destructive_2020,tinkham_introduction_2004}, and the sensitivity to the chemical potential, which requires delicate gating~\cite{potter_majorana_2011}. An important advancement came in the form of planar phase-biased Josephson junctions~\cite{hell_two-dimensional_2017,pientka_topological_2017,laeven_enhanced_2020} (also known as superconductor--normal--superconductor or SNS junctions).
There, time-reversal symmetry is broken by both an in-plane Zeeman field and superconducting phase bias, which define a two-dimensional parameter plane. Large regions of this parameter plane are topological, including, in principle, regions with weak magnetic fields. In the wire geometry one typically assumes that the level spacing between the transverse modes $d$ is larger than the induced superconducting gap~$\Delta$. In contrast, in the planar geometry $d \ll \Delta$, so that many modes participate in the formation of the topological state, and therefore the boundaries of the topological regions depend only weakly on the system's chemical potential~\cite{pientka_topological_2017,setiawan_topological_2019}.
Several experimental results indeed show the potential and versatility of this platform~\cite{ren_topological_2019,fornieri_evidence_2019,banerjee_signatures_2022,banerjee_local_2022,banerjee_control_2022}.
However, to get an appreciable topological gap, one still needs to apply significant magnetic fields.

Our goal in this manuscript is to induce topological superconductivity in a planar system, using only phase biasing and without applying any Zeeman field.
First steps in this direction were previously taken~\cite{fu_superconducting_2008,kotetes_topological_2015,riwar_multi-terminal_2016,melo_supercurrent-induced_2019,lesser_three-phase_2021,lesser_phase-induced_2021} (for a recent review see~\cite{lesser_majorana_2022}).
Here we provide a straightforward recipe based on simple principles which are sufficient to achieve this goal, and propose materials suitable for realizing our recipe.

Two key points allow us to accomplish this objective.
The first is the introduction of two phase differences by including three superconductors in our system. 
With these two phase differences, we show that phase winding can drive the system into a topological phase.
The second pivotal element is unequal Fermi velocities for the two spin branches in the direction transverse to the junction. 
The combination of these two ingredients can replace the external Zeeman field altogether.

\emph{Phase winding and unequal Fermi velocities.---}In one-dimensional superconducting systems where time-reversal symmetry is broken and translational invariance holds, a transition between trivial and topological phases occurs when there is a single gap closing at zero longitudinal momentum ($k_x=0$). For a single Josephson junction, the closure of the gap happens when a single pair of eigenstates of the Bogoliubov--de Gennes Hamiltonian crosses zero energy~\cite{kitaev_unpaired_2001} as we tune the phase bias across the junction.
Evidently, for such a single gap closing to occur, spin symmetry must be broken.  
Indeed, the application of the Zeeman field separates between the gap closing curves of the two spin branches in the parameter space, thereby separating trivial and topological regions~\cite{hell_two-dimensional_2017,pientka_topological_2017}.
We now show that when the two branches have different velocities, and the single Josephson junction is replaced by two junctions in series, the Zeeman field may be replaced by a second phase difference.

\begin{figure}[t]
    \centering
   \includegraphics[width=\linewidth]{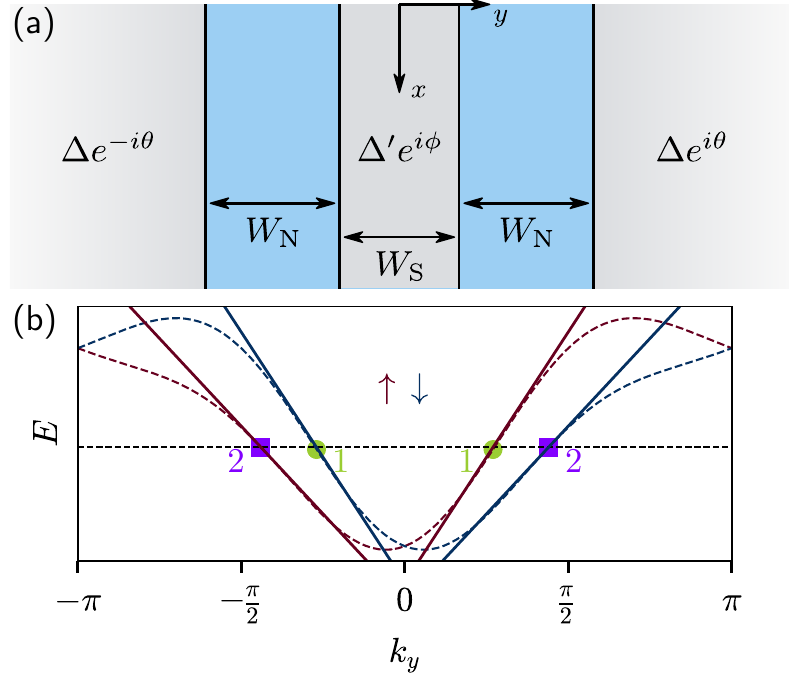}
    \caption{(a)~Phase-biased SNSNS junction (superconducting regions are gray, normal regions are blue). The two outer superconductors are semi-infinite in the transverse direction $y$, and their superconducting order parameter $\Delta$ is in general different than the one of the middle superconductor $\Delta^{\prime}$. At $k_x=0$ this reduces to a one-dimensional problem, whose zero-energy crossings correspond to topological phase transitions.
    (b)~Transverse spectrum of the junction (dashed lines), in a model including next-nearest neighbor hopping~\cite{SupplementalMaterial}. At the Fermi energy (dashed black line), there are four Fermi points. The Fermi velocities at the ``outer" branch (purple squares) and in the ``inner" branch (green dots) are not identical. Solid lines correspond to the linearized spectra.
    \label{fig:SNSNS_junction}}
\end{figure}

We first consider an SNS junction between two superconductors whose order parameters are $\Delta e^{\pm i\theta}$.
If the superconducting gap $\Delta$ is much smaller than the Fermi energy $E_{\rm F}$, then at $\theta=\pi/2$, where the phase difference across the junction is $\pi$, a double gap closing occurs, with four states at zero energy. 
Starting from this point, we introduce a third superconductor in the middle of the junction with an order parameter $\Delta^{\prime} e^{i\phi}$, see Fig.~\figref{fig:SNSNS_junction}{a}, and search for single gap-closing curves  in $\theta$--$\phi$ space. The regions between such curves are topological.

To this end, we construct a simple model describing an SNSNS geometry with two linearly dispersing branches $j=1,2$ with the normal-state Hamiltonian $H^\pm_{0,j}=\pm iv_j \partial_y$, where the $\pm$ indicate the two opposite directions of motion across the junction, see Fig.~\figref{fig:SNSNS_junction}{b}, and we set $\hbar=1$ (a similar system was studied in Refs.~\cite{hurd_superconducting_1995,chang_andreev-level_1997}).
As we study below, under certain conditions spin-orbit coupling may make the velocities $v_1$, $v_2$ unequal, which will be important in what follows.
At $k_x=0$, the system along the $y$ direction becomes one-dimensional, and finding the bound states involves a standard calculation whose details are given in Sec.~\ref{App:Exact} of the Supplemental Material~\cite{SupplementalMaterial}.
We find that for the $j$'th branch there is a single gap closing along a line in the $\theta$--$\phi$ plane defined by  
\begin{equation}\label{eq:zero_E_condition}
    \cos\theta + \tanh\left(\frac{W_{\rm S}\Delta^{\prime}}{v_j}\right)\cos\phi=0,
\end{equation}
where $W_{\rm S}$ is the width of the middle superconductor. 

Several aspects of Eq.~\eqref{eq:zero_E_condition} are noteworthy. First, we see that the position of the gap-closing transition within the $\theta$--$\phi$ plane is determined by the dimensionless ratio $W_{\rm S}/\xi_{j}$, where $\xi_{j} = v_j/\Delta^{\prime}$ is the coherence length of the middle superconductor for branch $j$. When the middle superconductor is absent ($W_{\rm S}=0$), the transition occurs for both branches at $\theta=\pi/2+\pi n$ (here $n$ is an integer). When the middle superconductor is very wide, the system may be seen as two disconnected junctions, and a gap closing takes place when the phase difference across \emph{one} junction is $\phi\pm\theta=\pi+2\pi n$.
In between, the position of the gap closing creates a curve in the $\theta$--$\phi$ plane. When the velocities in the middle superconductor $v_1$, $v_2$ are unequal, the gap-closing curves  for the two branches are different [see Eq.~\eqref{eq:zero_E_condition}], and constitute topological class D phase transitions. The topological region they define is maximized when, without loss of generality, $\xi_1\ll W_{\rm S}$ and $\xi_2\gg W_{\rm S}$  (see Fig.~\ref{fig:exact_PD_SNSNS}, and the discussion in Sec.~\ref{App:opt} of the Supplemental Material~\cite{SupplementalMaterial}).
Interestingly, as we show in Sec.~\ref{App:Exact} of the Supplemental Material~\cite{SupplementalMaterial}, Eq.~\eqref{eq:zero_E_condition} is independent of the width and velocities of the normal regions. This independence has important practical consequences, since generally a narrow normal part leads to a relatively large gap for the Andreev states away from the transitions. In previous proposals~\cite{lesser_topological_2020, hell_two-dimensional_2017,pientka_topological_2017} other considerations did not allow for arbitrarily narrow normal parts. 

\begin{figure}[t]
    \centering
   \includegraphics[width=\linewidth]{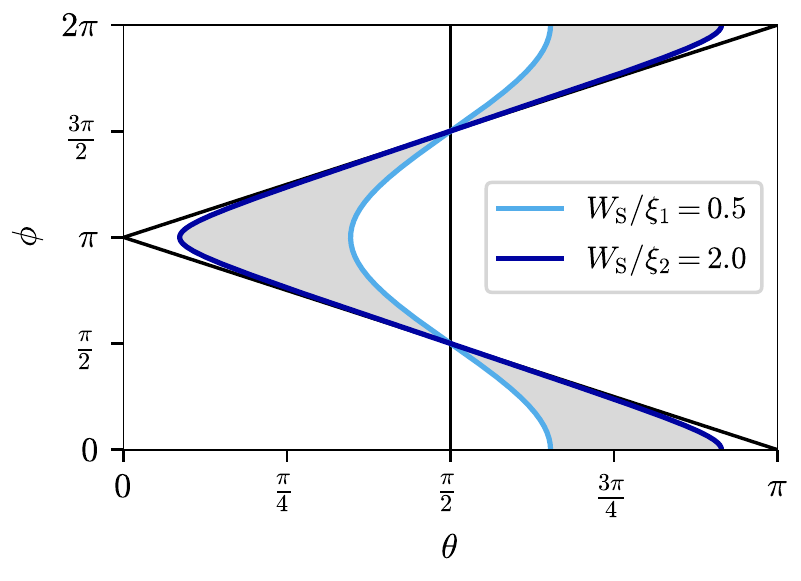}
    \caption{Phase diagram for the SNSNS geometry of Fig.~\ref{fig:SNSNS_junction}, derived from Eq.~\eqref{eq:zero_E_condition}. The two values $\xi_{j}=v_{j}/\Delta^{\prime}$ originate from the unequal transverse Fermi velocities, $v_1 \neq v_2$. The light and dark blue curves correspond to zero-energy crossings at $k_x=0$ for $W_{\rm S}/\xi_{1}=0.5$ and $W_{\rm S}/\xi_{2}=2$, respectively. Since each of these crossings is non-degenerate, it corresponds to a topological phase transition, and therefore the area between the curves (shaded) harbors a topological superconducting state. The black lines define the region where a vortex is present.
    \label{fig:exact_PD_SNSNS}}
\end{figure}

Although our analysis assumed the absence of disorder, the emerging picture suggests a qualitative criterion for its effect. As we saw, in order to obtain a well-separated single gap closing, one branch should traverse the entire middle superconductor, while the other should be reflected back from that superconductor. To this end, it is required that disorder will be weak enough such that no significant inter-branch scattering would occur on a scale of $W_{\rm S}$. In addition we require that smooth potential fluctuations are sufficiently weak to keep the velocities unequal.

Our analysis, which focused on the case of two branches, may easily be generalized to situations where there are more branches. This could happen, for example, due to having more than one subband in the $z$ direction, or due to having more than two bands with states at the Fermi energy. With many branches there would be many single gap closings, and therefore many transitions between trivial and topological regions in the $\theta$--$\phi$ plane. A well-separated single gap closing requires a Fermi velocity that is significantly different from the other Fermi velocities. 

Remarkably, for all values of $W_{\rm S}/\xi$ the curves defined by Eq.~\eqref{eq:zero_E_condition} and the topological regions they induce correspond to phase configurations in which the phase winds by $2\pi$ (in agreement with Ref.~\cite{van_heck_single_2014}).
As shown previously~\cite{lesser_three-phase_2021}, a $2\pi$ phase winding occurs when $g=(\cos\theta+\cos\phi)\cos\theta<0$~\cite{SupplementalMaterial}. By Eq.~\eqref{eq:zero_E_condition}, 
\begin{equation}
    g = -\tanh\left(\frac{W_{\rm S}}{\xi_j}\right)\left[ 1 - \tanh\left(\frac{W_{\rm S}}{\xi_j}\right) \right]\cos^2\phi.
    \label{eq:winding}
\end{equation}
Since $0<\tanh\left(W_{\rm S}/\xi_{j}\right)<1$, Eq.~\eqref{eq:winding} always implies a phase winding. 

\emph{Obtaining unequal Fermi velocities for the two spin branches.---}Following the above observations, we now turn to search for setups in which the Fermi velocities for the two spin branches are unequal. 
We begin by discussing several material platforms where the band structure naturally has such an imbalance, and continue by discussing how it may be artificially created.

Monolayers of  transition metal dichalcogenides (TMDs)~\cite{wang_electronics_2012,xiao_coupled_2012,qian_quantum_2014,manzeli_2d_2017,wu_observation_2018,lee_topological_2019} constitute a platform that is particularly suitable for our purposes, for two reasons. First, due to the strong spin-orbit coupling in TMDs, it is quite easy to find directions along which the Fermi velocities distinguish between the spin branches.
In ${\rm TaS}_2$, for example, we found a velocity imbalance as large as $v_{<}/v_{>}=0.7$ using an effective six-band Hamiltonian~\cite{margalit_theory_2021} (here $v_>$ is the larger velocity of the two and $v_<$ is the smaller one).
Moreover, we used a well-studied six-band tight-binding model~\cite{liu_three-band_2013} of $MX_2$ ($M={\rm Mo, W}$; $X={\rm S, Se, Te}$) to show that these TMDs also support similar velocity imbalances.

Second, the SOC in TMDs is of the Ising type, in which the spins are polarized in the out-of-plane direction, with the sign of polarization depending on the momentum.
This property endows TMDs their giant critical field in the superconducting state, but it also makes them unsuitable for the setups of Refs.~\cite{hell_two-dimensional_2017,pientka_topological_2017}, where the SOC suppresses the spins' sensitivity to a Zeeman field. Since our scheme makes no use of a Zeeman field, this difficulty is alleviated. 
The remarkable combination of intrinsic gate-controlled superconductivity and unequal Fermi velocities in monolayers of ${\rm WTe}_2$, for example,  could open the door to topological superconductivity in a single material system, without the need to proximity couple it to an external superconductor.

Beyond TMDs, calculations for quasi-one dimensional wires defined by gates operating on HgTe quantum wells~\cite{reuther_gate-defined_2013} show a similar velocity imbalance~\cite{SupplementalMaterial}. 
Given the velocity imbalance, our theory provides a practical guide to designing the device geometry to optimize the stability of the topological phase~\cite{SupplementalMaterial}.

\emph{Inducing unequal Fermi velocities: general considerations.---}To analyze possible sources for velocity imbalance, it is instructive to start from one-dimensional systems. 
In a 1D ring constrained to have only nearest-neighbor hopping, the most general spin-orbit coupling leads to the dispersion $E_\pm(k)$ of the two spin branches being rigidly shifted along the $k$-axis, $E_{\pm}\left(k\right)=E\left(k\pm k_{\rm SO}\right)$~\cite{meir_universal_1989}. This may be understood by realizing that spin-orbit coupling introduces a spin-dependent Aharonov--Casher (AC) flux into the loop defined by the ring~\cite{aharonov_topological_1984}. Then, the Fermi velocities (defined by $v_\pm=\partial E_\pm(k)/\partial k$) of the two spin branches are necessarily identical. The introduction of longer-range hopping, such as between next-nearest neighbors, expands the number of loops threaded by AC fluxes. Spin-orbit coupling then has a richer effect on the spectrum, which in general leads to unequal Fermi velocities. 

\begin{figure}[t]
    \centering
   \includegraphics[width=\linewidth]{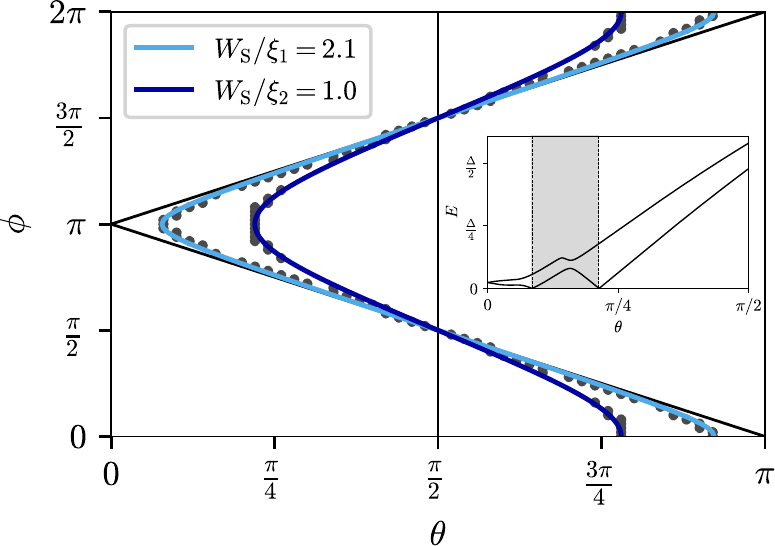}
    \caption{Tight-binding simulation of the SNSNS junction of Fig.~\ref{fig:SNSNS_junction} at $k_x=0$, showing the phase diagram as a function of the two phases $\theta$ and $\phi$. The gray dots are the locations of the zero-energy crossings as calculated by the tight-binding simulation. The light and dark blue curves are fits of these points according to the analytical formula, Eq.~\eqref{eq:zero_E_condition}. We observe good agreement between the formula and the simulation (cf. Fig.~\ref{fig:exact_PD_SNSNS}), which also enables us to extract the two effective coherence lengths $\xi_{1}$, $\xi_{2}$.
    Inset: bound-state spectrum as a function of $\theta$, the phase of the middle superconductor, for $\phi=\pi$. The zero-energy crossings split along $\theta$, rather than being degenerate. The resulting topological region is marked in gray.
    The simulation was performed according to Eq.~\eqref{eq:H2D} with the parameters  $t_1=t_2=1$, $\lambda_{1,y}=\lambda_{2,x}=1.2$ (all other $\lambda_{i,\alpha}=0$), $\mu=3$, $\Delta=\Delta^{\prime}=0.1$, $W_{\rm N}=0$, $W_{\rm S}=80$. The left and right superconductors are each 200 sites wide.
    \label{fig:TB}}
\end{figure}

As we now show, in two dimensions nearest-neighbor hopping is sufficient to generate unequal Fermi velocities of the two branches, when projected on to a certain direction. The most general 2D band Hamiltonian of a monoatomic unit cell which is time-reversal symmetric and limited to nearest-neighbor hopping is 
\begin{equation}
    H_{\rm 2D}({\bf k})=-\sum_it_i\cos\left({\bf k}\cdot{\bf a}_i\right) - \sum_{i,\alpha}\lambda^i_{\alpha}\sin\left({\bf k}\cdot{\bf a}_i\right)\sigma_\alpha,
    \label{eq:H2D}
\end{equation}
where ${\bf a}_i$ ($i=1,2$) are the lattice's unit vectors, and $\alpha=x,y,z$ are Pauli matrix indices. We are interested in the dispersion and the velocities for $k_x=0$. The $x$ direction lies along the junction, and the orientations of the vectors ${\bf a}_i$ are left for tuning. If we choose the orientation of the lattice such that $a_{1,y}/a_{2,y}=n$, the 2D Hamiltonian $H_{\rm 2D}(k_x=0,k_y)$ is identical to a 1D Hamiltonian with lattice constant $a_{1,y}$, and hopping amplitude to distances of $a_{1,y}$ and $na_{1,y}$, i.e., further neighbor hopping. For $n=0,1$ the problem maps onto the 1D ring with only nearest-neighbor coupling, with identical Fermi velocities to both branches, while for $n\ne 0, 1$ the velocities are generically unequal.
This mapping may easily be generalized to the case $a_{1,y}/a_{2,y}=n/m$ (with integer $m,n$ and $m>0$) and to 2D Hamiltonians that include long-range hopping amplitudes. Tight-binding calculations supporting this analysis are shown in Fig.~\ref{fig:TB}.

We note, however, that for the angle dependence of the velocity to be manifest, we typically need to go beyond the second-order-in-$k$ expansion of Eq.~\eqref{eq:H2D}. At that order, the band structure for each angle is parabolic, and spin-orbit coupling results in a mere rigid shift of the parabolas. The parabolic approximation ceases to hold at large electronic densities, i.e., inter-electron distance that is comparable to the lattice constant. Such densities are uncommon  in semiconducting heterostructure-based two-dimensional electronic systems. Two different mechanisms may lead to unequal Fermi velocities of the two spin branches even at low densities. The first requires more than one subband in the $z$ direction of the 2D electron gas with spin-orbit coupling strength depending on $z$ ~\cite{lesser_three-phase_2021} and the second requires a periodic potential~\cite{lesser_phase-induced_2021}.

The first mechanism may be understood by considering a heterostructure defined by a confining potential $V(z)$. In the absence of spin-orbit coupling, the confining potential defines spin-degenerate subbands in the $z$ direction, with the first two characterized by confinement energies $0,\delta E$ and by the real wavefunctions $\chi_1(z),\chi_2(z)$. The dispersion of each subband is quadratic. At low densities all higher subbands may be neglected, and we can project the spin-orbit coupling to the subspace defined by the two lowest subbands. Rashba SOC takes the form $H_{\rm R}=\alpha(z){\bf k}\times {\boldsymbol\sigma}\cdot {\hat z}$.  Taking for simplicity $k_x=0$ and projecting $H_{\rm R}$ to the subspace of the two lowest subbands we get, due to the spin, a $4\times 4$ Hamiltonian,
\begin{equation}\label{eq:alphaz}
    H=\begin{pmatrix} \frac{k_y^2}{2m^*}& 0 \\ 0 & \frac{k_y^2}{2m^*} + \delta E \end{pmatrix}\sigma_0 +
    \begin{pmatrix}
    \alpha_{11}& \alpha_{12} \\
    \alpha_{12}& \alpha_{22} 
    \end{pmatrix}k_y\sigma_x,
\end{equation}
in which $m^*$ is the effective electron mass and $\alpha_{ij}\equiv \int dz \chi_i(z)\alpha(z)\chi_j(z)$. For $z$-independent $\alpha$ we have $\alpha_{12}=0$. The subbands are then decoupled, and they are spin split to two shifted parabolas with equal velocities. When $\alpha$ depends on $z$ the two subbands are coupled, and the velocities become unequal~\cite{tosi_spin-orbit_2019}. Details are given in Sec.~\ref{App:Twobands} of the Supplemental Material~\cite{SupplementalMaterial}. 

A second mechanism for inducing unequal Fermi velocities is combining Rashba SOC with a periodic potential \emph{along} the junction (in the $x$ direction).
A potential with wave vector $q$ mixes the states at $k_x=0$, which do not experience SOC, with states at $k_x=\pm q$, where SOC cannot be gauged away.
Generically this leads to unequal Fermi velocities of the transverse modes.
As a minimal model which demonstrates this possibility, we consider three superconducting pads deposited on top of a spin-orbit-coupled 2D electron gas.
We apply a periodic modulation to the chemical potential of the middle superconductor, $\mu(x)=\mu_0 + \delta\mu\cos\left(2\pi qx\right)$.
Using a tight-binding discretization, we numerically calculate the spectrum and the topological invariant~\cite{wimmer_algorithm_2012} for some integer values of $q$, and find a topological phase transition~\cite{SupplementalMaterial} (see Sec.~\ref{App:periodic_potential} of the Supplemental Material~\cite{SupplementalMaterial}).

\emph{Outlook.---}Our work suggests a purely phase-controlled setup that induces topological superconductivity in an SNSNS devices, composed of two Josephson junctions in series. 
Our mechanism  requires two conditions: superconducting phase winding and unequal Fermi velocities for the two electron branches. We find that the winding of the phase, which guarantees
a net current flow through the two junctions, is a necessary condition for the topological state to form.

Unequal Fermi velocities in our setup are a consequence of spin-orbit coupling. 
We pointed out several examples of materials in which band-structure calculations predict unequal Fermi velocities for the two spin branches. Then, by studying microscopic models, we showed how unequal Fermi velocities may be generated at low densities, characteristic of semiconducting heterostructures. 
Remarkably, a common thread to the models we studied here, and those of earlier suggestions~\cite{lesser_three-phase_2021,lesser_phase-induced_2021}, is the essential role played by closed loops traversed by electrons and holes in which a spin-dependent Aharonov--Casher phase is accumulated. 
In the full-shell models~\cite{vaitiekenas_flux-induced_2020,lesser_three-phase_2021} the closed trajectories are along the azimuthal cylinder direction; in the periodic potential case these are in-plane closed orbits; and in the tight-binding model closed trajectories along a triangle in a basic unit cell will accumulate a nontrivial Aharonov--Casher phase.
It seems that the combination of a discrete vortex in the three superconductors, superimposed with an Aharonov--Casher phase in closed trajectories leading to unequal Fermi velocities, is necessary to replace the Zeeman field in phase-only recipes for topological superconductivity.

The ubiquity of materials and engineered devices having unequal Fermi velocities makes our proposal within reach of current experiments. 
The elimination of any applied Zeeman field should greatly aid in achieving reliable experimental results.
Furthermore, the inherent periodicity of the phases will help distinguish the topological effect from trivial ones, and reliably map out the phase diagram (see Fig.~\ref{fig:extended_PD} of the Supplemental Material~\cite{SupplementalMaterial}).

The code used for simulating the models and generating the plots in this study is available at \url{http://dx.doi.org/10.5281/zenodo.6645458}.

\emph{Acknowledgment.---}We are grateful to Charles Kane, Charles Marcus and Amir Yacoby for insightful discussions, and to Gilad Margalit for advice on the band structure of TMDs.
The work was supported by the European Union's Horizon 2020 research and innovation programme (Grant Agreement LEGOTOP No. 788715), the DFG (CRC/Transregio 183, EI 519/7-1), ISF Quantum Science and Technology (2074/19), the BSF and NSF (2018643).

\bibliography{library,library_supp}


\clearpage
\setcounter{secnumdepth}{2}
\onecolumngrid

\begin{center}
\Large{\textbf{Supplemental Material}}
\end{center}

\setcounter{equation}{0}
\renewcommand{\theequation}{S\arabic{equation}}
\setcounter{figure}{0}
\renewcommand{\thefigure}{S\arabic{figure}}
\setcounter{section}{0}
\renewcommand{\thesection}{S\Roman{section}}

\section{Exact solution of the SNSNS junction}

\label{App:Exact}
Here we derive the bound-state spectrum of a one-dimensional SNSNS junction.
This should be understood as the $k_x=0$ solution of the quasi-one-dimensional system depicted in Fig.~\ref{fig:SNSNS_junction} of the main text.

We begin by obtaining the spectrum of linearly dispersing electrons in the presence of a semi-infinite superconductor.
The Bogoliubov--de Gennes Hamiltonian is 
\begin{equation}
    H=\begin{pmatrix}-iv\partial_{y} & \Delta e^{i\gamma}\\
\Delta e^{-i\gamma} & iv\partial_{y}
\end{pmatrix},
\end{equation}
where $v$ is the Fermi velocity, $\Delta$ is the pair potential, and $\gamma$ is the superconducting phase. 
We are interested in finding sub-gap states, which are evanescent waves.
To this end, we assume 
\begin{equation}
    \begin{pmatrix}\psi_{e}\left(y\right)\\
\psi_{h}\left(y\right)
\end{pmatrix}=\begin{pmatrix}\chi_{e}\\
\chi_{h}
\end{pmatrix}e^{iky},
\end{equation}
and substitute into the Hamiltonian:
\begin{equation}
    H=\begin{pmatrix}vk & \Delta e^{i\gamma}\\
\Delta e^{-i\gamma} & -vk
\end{pmatrix}.
\end{equation}
The eigen-energies are $E=\pm\sqrt{v^{2}k^{2}+\Delta^{2}}$, so that $vk=\pm i\sqrt{\Delta^{2}-E^{2}}$, and the sign depends on whether the SC is semi-infinite towards $y\to\infty$ or $y\to-\infty$.
It is convenient to denote
\begin{equation}
    E=\Delta\cos\alpha\ \Rightarrow k=\pm\frac{i\Delta\sin\alpha}{v}.
\end{equation}
The corresponding eigenstates can be parameterized as
\begin{equation}\label{seq:S_wavefunction}
    \begin{pmatrix}\chi_{e}\\
\chi_{h}
\end{pmatrix}=\frac{1}{\sqrt{2}}\begin{pmatrix}1\\
e^{i\beta}
\end{pmatrix},
\end{equation}
and the phase $\beta$ satisfies 
\begin{equation}\label{seq:beta_general}
    \alpha+\beta+\gamma=2\pi n,\quad(n\in\mathbb{Z}).
\end{equation}

\begin{figure}[b]
    \centering
   \includegraphics[width=0.7\linewidth]{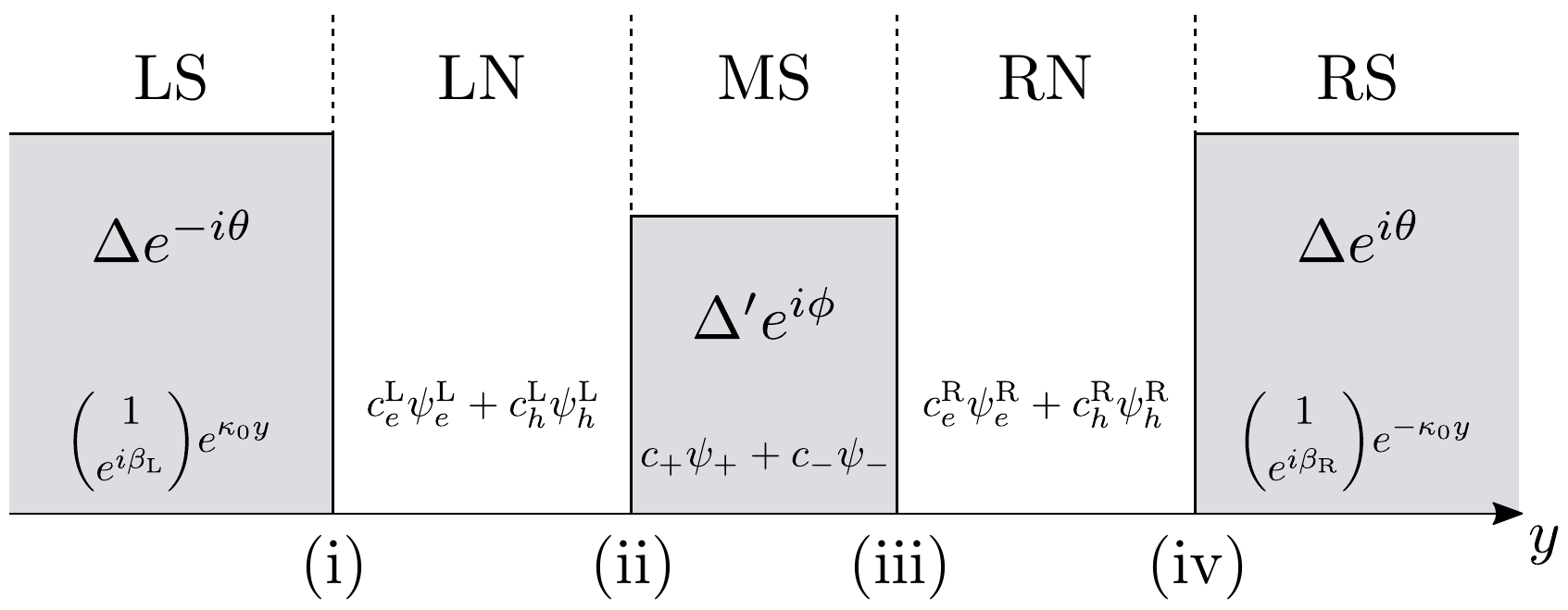}
    \caption{One-dimensional SNSNS junction. with the different regions (left/middle/right, superconductor/normal) marked. The interfaces between the regions are marked by (i)--(iv). The wavefunctions in each region are indicated.
    \label{sfig:SNSNS_schematic}}
\end{figure}

The system of interest contains five regions, as illustrated in Fig.~\ref{sfig:SNSNS_schematic}.
We label the regions by two letters, L/M/R (for left/right/middle) S/N (for superconductor/normal).
The pair potential and velocity in each region are given by: 
\begin{center}
\begin{tabular}{|c|c|c|c|}
    \hline
     Region & $y$ range & $\Delta(y)$ & $v(y)$ \\ [0.5ex]
 \hline\hline
     LS & $y<-W_{\text{N}}-\frac{W_{\text{S}}}{2}$ & $\Delta e^{-i\theta}$ & $v_{\rm S}$ \\ \hline
     LN & $-W_{\text{N}}-\frac{W_{\text{S}}}{2}<y<-\frac{W_{\text{S}}}{2}$ & $0$ & $v_{\rm N}$ \\ \hline
     MS & $-\frac{W_{\text{S}}}{2}<y<\frac{W_{\text{S}}}{2}$ & $\Delta^{\prime} e^{i\phi}$ & $v_{\rm S}$ \\ \hline
     RN & $\frac{W_{\text{S}}}{2}<y<W_{\text{N}}+\frac{W_{\text{S}}}{2}$ & $0$ & $v_{\rm N}$ \\ \hline
     RS & $y>W_{\text{N}}+\frac{W_{\text{S}}}{2}$ & $\Delta e^{i\theta}$ & $v_{\rm S}$ \\ \hline
\end{tabular}
\end{center}

The wavefunctions in the normal regions are just plane waves.
The wavefunctions in the superconducting regions have been found in Eq.~\eqref{seq:S_wavefunction}; in the middle superconductor the wavefunction may either propagate or decay, as it is finite. 
We may therefore write the wavefunctions as 
\begin{center}
\begin{tabular}{|c|c|}
\hline 
LS & $\begin{pmatrix}1\\
e^{i\beta_{\text{L}}}
\end{pmatrix}e^{-\kappa_{0}\left(y+\frac{W_{\text{S}}}{2}+W_{\text{N}}\right)}$\\
\hline 
LN & $\psi_{e}^{\text{L}}=\begin{pmatrix}e^{i\frac{E}{v_{\text{N}}}\left(y+\frac{W_{\text{S}}}{2}+W_{\text{N}}\right)}\\
0
\end{pmatrix},\ \psi_{h}^{\text{L}}=\begin{pmatrix}0\\
e^{-i\frac{E}{v_{\text{N}}}\left(y+\frac{W_{\text{S}}}{2}+W_{\text{N}}\right)}
\end{pmatrix}$\tabularnewline
\hline 
MS & $\psi_{+}=\begin{pmatrix}1\\
e^{i\beta_{+}}
\end{pmatrix}e^{\kappa y},\ \psi_{-}=\begin{pmatrix}1\\
e^{i\beta_{-}}
\end{pmatrix}e^{-\kappa y}$\tabularnewline
\hline 
RN & $\psi_{e}^{\text{R}}=\begin{pmatrix}e^{i\frac{E}{v_{\text{N}}}\left(y-\frac{W_{\text{S}}}{2}\right)}\\
0
\end{pmatrix},\ \psi_{h}^{\text{R}}=\begin{pmatrix}0\\
e^{-i\frac{E}{v_{\text{N}}}\left(y-\frac{W_{\text{S}}}{2}\right)}
\end{pmatrix}$\tabularnewline
\hline 
RS & $\begin{pmatrix}1\\
e^{i\beta_{\text{R}}}
\end{pmatrix}e^{-\kappa_{0}\left(y-\frac{W_{\text{S}}}{2}-W_{\text{N}}\right)}$\tabularnewline
\hline 
\end{tabular}
\end{center}

Here $\kappa_{0}=\sqrt{\Delta^{2}-E^{2}}/v_{\text{S}}$, $\kappa=\sqrt{\Delta^{\prime2}-E^{2}}/v_{\text{S}}$ (notice that $\kappa$ may take real or imaginary values). The $\beta$ values are, in accordance with Eq.~\eqref{seq:beta_general}, given by 
\begin{equation}\label{seq:beta_all}
    \begin{aligned}\beta_{\text{L}} & =\theta+\cos^{-1}\frac{E}{\Delta}\\
\beta_{\text{R}} & =-\theta-\cos^{-1}\frac{E}{\Delta}=-\beta_{\text{L}}\\
\beta_{+} & =-\phi+\cos^{-1}\frac{E}{\Delta^{\prime}}\\
\beta_{-} & =-\phi-\cos^{-1}\frac{E}{\Delta^{\prime}}.
\end{aligned}
\end{equation}
We now match the wavefunctions at the four interfaces (notice that there is no continuity condition on the derivative of the wavefunction, since the Hamiltonian is of first order).

\begin{enumerate}[(i)]
    \item \textbf{LS--LN.} We write the wavefunction at LN as $c_{e}^{\text{L}}\psi_{e}^{\text{L}}+c_{h}^{\text{L}}\psi_{h}^{\text{L}}$.
    Continuity at $y=-W_{\text{N}}-W_{\text{S}}/2$ implies
    \begin{equation}
    \begin{pmatrix}1\\
        e^{i\beta_{\text{L}}}
        \end{pmatrix}=\begin{pmatrix}c_{e}^{\text{L}}\\
        c_{h}^{\text{L}}
        \end{pmatrix}.
    \end{equation}
    
    \item \textbf{LN--MS.} We write the wavefunction at MS as $c_{+}\psi_{+}+c_{-}\psi_{-}$. Continuity at $y=-W_{\text{S}}/2$ implies 
    \begin{equation}
    \begin{pmatrix}c_{e}^{\text{L}}e^{iEW_{\text{N}}/v_{\text{N}}}\\
c_{e}^{\text{L}}e^{-iEW_{\text{N}}/v_{\text{N}}}
\end{pmatrix}=\begin{pmatrix}c_{+}e^{-\kappa W_{\text{S}}/2}+c_{-}e^{\kappa W_{\text{S}}/2}\\
c_{+}e^{i\beta_{+}}e^{-\kappa W_{\text{S}}/2}+c_{-}e^{i\beta_{-}}e^{\kappa W_{\text{S}}/2}
\end{pmatrix}.    
    \end{equation}
    The coefficients $c_{+},c_{-}$ may be expressed as
    \begin{equation}
        \begin{pmatrix}c_{+}\\
c_{-}
\end{pmatrix}=\begin{pmatrix}e^{-\frac{\kappa W_{\text{S}}}{2}} & e^{\frac{\kappa W_{\text{S}}}{2}}\\
e^{i\beta_{+}}e^{-\frac{\kappa W_{\text{S}}}{2}} & e^{i\beta_{-}}e^{\frac{\kappa W_{\text{S}}}{2}}
\end{pmatrix}^{-1}\begin{pmatrix}e^{i\frac{EW_{\text{N}}}{v_{\text{N}}}}\\
 & e^{-i\frac{EW_{\text{N}}}{v_{\text{N}}}}
\end{pmatrix}\begin{pmatrix}1\\
e^{i\beta_{\text{L}}}
\end{pmatrix}.
    \end{equation}
    
    \item \textbf{MS--RN.} We write the wavefunction at RN as $c_{e}^{\text{R}}\psi_{e}^{\text{R}}+c_{h}^{\text{R}}\psi_{h}^{\text{R}}$. Continuity at $y=W_{\text{S}}/2$ implies
    \begin{equation}
\begin{pmatrix}c_{+}e^{\kappa W_{\text{S}}/2}+c_{-}e^{-\kappa W_{\text{S}}/2}\\
c_{+}e^{i\beta_{+}}e^{\kappa W_{\text{S}}/2}+c_{-}e^{i\beta_{-}}e^{-\kappa W_{\text{S}}/2}
\end{pmatrix}=\begin{pmatrix}c_{e}^{\text{R}}\\
c_{h}^{\text{R}}
\end{pmatrix}.
    \end{equation}
    The coefficients $c_{e}^{\text{R}},c_{e}^{\text{L}}$ may be expressed as
    \begin{equation}
    \begin{aligned}\begin{pmatrix}c_{e}^{\text{R}}\\
c_{h}^{\text{R}}
\end{pmatrix} & =\begin{pmatrix}e^{\frac{\kappa W_{\text{S}}}{2}} & e^{-\frac{\kappa W_{\text{S}}}{2}}\\
e^{i\beta_{+}}e^{\frac{\kappa W_{\text{S}}}{2}} & e^{i\beta_{-}}e^{-\frac{\kappa W_{\text{S}}}{2}}
\end{pmatrix}\begin{pmatrix}e^{-\frac{\kappa W_{\text{S}}}{2}} & e^{\frac{\kappa W_{\text{S}}}{2}}\\
e^{i\beta_{+}}e^{-\frac{\kappa W_{\text{S}}}{2}} & e^{i\beta_{-}}e^{\frac{\kappa W_{\text{S}}}{2}}
\end{pmatrix}^{-1}\\
 & \times\begin{pmatrix}e^{i\frac{EW_{\text{N}}}{v_{\text{N}}}}\\
 & e^{-i\frac{EW_{\text{N}}}{v_{\text{N}}}}
\end{pmatrix}\begin{pmatrix}1\\
e^{i\beta_{\text{L}}}
\end{pmatrix}.
\end{aligned}
    \end{equation}
    
    \item \textbf{RN--RS.} Finally, we write the wavefunction at RS as some coefficient $c_{\text{S}}$ multiplied by the unnormalized wavefunction, and impose continuity at $y=W_{\text{N}}+W_{\text{S}}/2$:
    \begin{equation}
        \begin{pmatrix}c_{e}^{\text{R}}e^{i\frac{EW_{\text{N}}}{v_{\text{N}}}}\\
c_{h}^{\text{R}}e^{-i\frac{EW_{\text{N}}}{v_{\text{N}}}}
\end{pmatrix}=c_{\text{S}}\begin{pmatrix}1\\
e^{i\beta_{\text{R}}}
\end{pmatrix}.
    \end{equation}
    In matrix form, this equation reads 
    \begin{equation}
        \begin{aligned}c_{\text{S}}\begin{pmatrix}1\\
e^{i\beta_{\text{R}}}
\end{pmatrix} & =\begin{pmatrix}e^{i\frac{EW_{\text{N}}}{v_{\text{N}}}}\\
 & e^{-i\frac{EW_{\text{N}}}{v_{\text{N}}}}
\end{pmatrix}\begin{pmatrix}e^{\frac{\kappa W_{\text{S}}}{2}} & e^{-\frac{\kappa W_{\text{S}}}{2}}\\
e^{i\beta_{+}}e^{\frac{\kappa W_{\text{S}}}{2}} & e^{i\beta_{-}}e^{-\frac{\kappa W_{\text{S}}}{2}}
\end{pmatrix}\\
 & \times\begin{pmatrix}e^{-\frac{\kappa W_{\text{S}}}{2}} & e^{\frac{\kappa W_{\text{S}}}{2}}\\
e^{i\beta_{+}}e^{-\frac{\kappa W_{\text{S}}}{2}} & e^{i\beta_{-}}e^{\frac{\kappa W_{\text{S}}}{2}}
\end{pmatrix}^{-1}\begin{pmatrix}e^{i\frac{EW_{\text{N}}}{v_{\text{N}}}}\\
 & e^{-i\frac{EW_{\text{N}}}{v_{\text{N}}}}
\end{pmatrix}\begin{pmatrix}1\\
e^{i\beta_{\text{L}}}
\end{pmatrix}.
\end{aligned}
    \end{equation}
\end{enumerate}
We now have an equation of the form
\begin{equation}
    c_{\text{S}}\begin{pmatrix}1\\
e^{i\beta_{\text{R}}}
\end{pmatrix}=M\begin{pmatrix}1\\
e^{i\beta_{\text{L}}}
\end{pmatrix},
\end{equation}
where $M$ plays the role of a transfer matrix. In order for this equation to hold, we eliminate $c_{\text{S}}$ by demanding that the ratio between the components of the vectors is identical:
\begin{equation}
    e^{i\beta_{\text{R}}}=\frac{M_{21}+M_{22}e^{i\beta_{\text{L}}}}{M_{11}+M_{12}e^{i\beta_{\text{L}}}}.
\end{equation}
Plugging all the previously obtained expression, we find
\begin{equation}
    \frac{e^{i(\alpha+\theta-\phi)}\sinh\left(\kappa W_{{\rm S}}\right)-e^{2i\left(\alpha+\theta-\frac{EW_{{\rm N}}}{v_{{\rm N}}}\right)}\sinh\left(\kappa W_{{\rm S}}+i\alpha^{\prime}\right)}{e^{\frac{2iEW_{{\rm N}}}{v_{{\rm N}}}}\sinh\left(\kappa W_{{\rm S}}-i\alpha^{\prime}\right)-e^{i(\alpha+\theta+\phi)}\sinh\left(\kappa W_{{\rm S}}\right)}=1,
\end{equation}
where $\alpha=\cos^{-1}\left(E/\Delta\right)$, $\alpha^{\prime}=\cos^{-1}\left(E/\Delta^{\prime}\right)$.

Let us now find the conditions for a zero-energy solution. We set $E=0$, which implies $\alpha=\alpha^{\prime}=\pi/2$, resulting in 
\begin{equation}
    \frac{e^{i(\theta-\phi)}\sinh\left(\kappa W_{\rm S}\right)+e^{2i\theta}\cosh\left(\kappa W_{\rm S}\right)}{\cosh\left(\kappa W_{\rm S}\right)+e^{i(\theta+\phi)}\sinh\left(\kappa W_{\rm S}\right)}=-1.
\end{equation}
Let us solve this equation:
\begin{equation}
\begin{gathered}e^{i(\theta-\phi)}\sinh\left(\kappa W_{\rm S}\right)+e^{2i\theta}\cosh\left(\kappa W_{\rm S}\right)=-\cosh\left(\kappa W_{\rm S}\right)-e^{i(\theta+\phi)}\sinh\left(\kappa W_{\rm S}\right)\\
\cosh\left(\kappa W_{\rm S}\right)\left(1+e^{2i\theta}\right)=-\sinh\left(\kappa W_{\rm S}\right)e^{i\theta}\left(e^{i\phi}+e^{-i\phi}\right)\\
\cosh\left(\kappa W_{\rm S}\right)e^{i\theta}\cos\left(\theta\right)=-\sinh\left(\kappa W_{\rm S}\right)e^{i\theta}\cos\left(\phi\right)\\
\cos\left(\theta\right)=-\tanh\left(\frac{W_{\rm S}}{\xi}\right)\cos\left(\phi\right).
\end{gathered}
\end{equation}
This final equation is identical Eq.~\eqref{eq:zero_E_condition} of the main text, with $\xi=v_{\rm S}/\Delta^{\prime}$.

\begin{figure}[t]
    \centering
   \includegraphics[width=\linewidth]{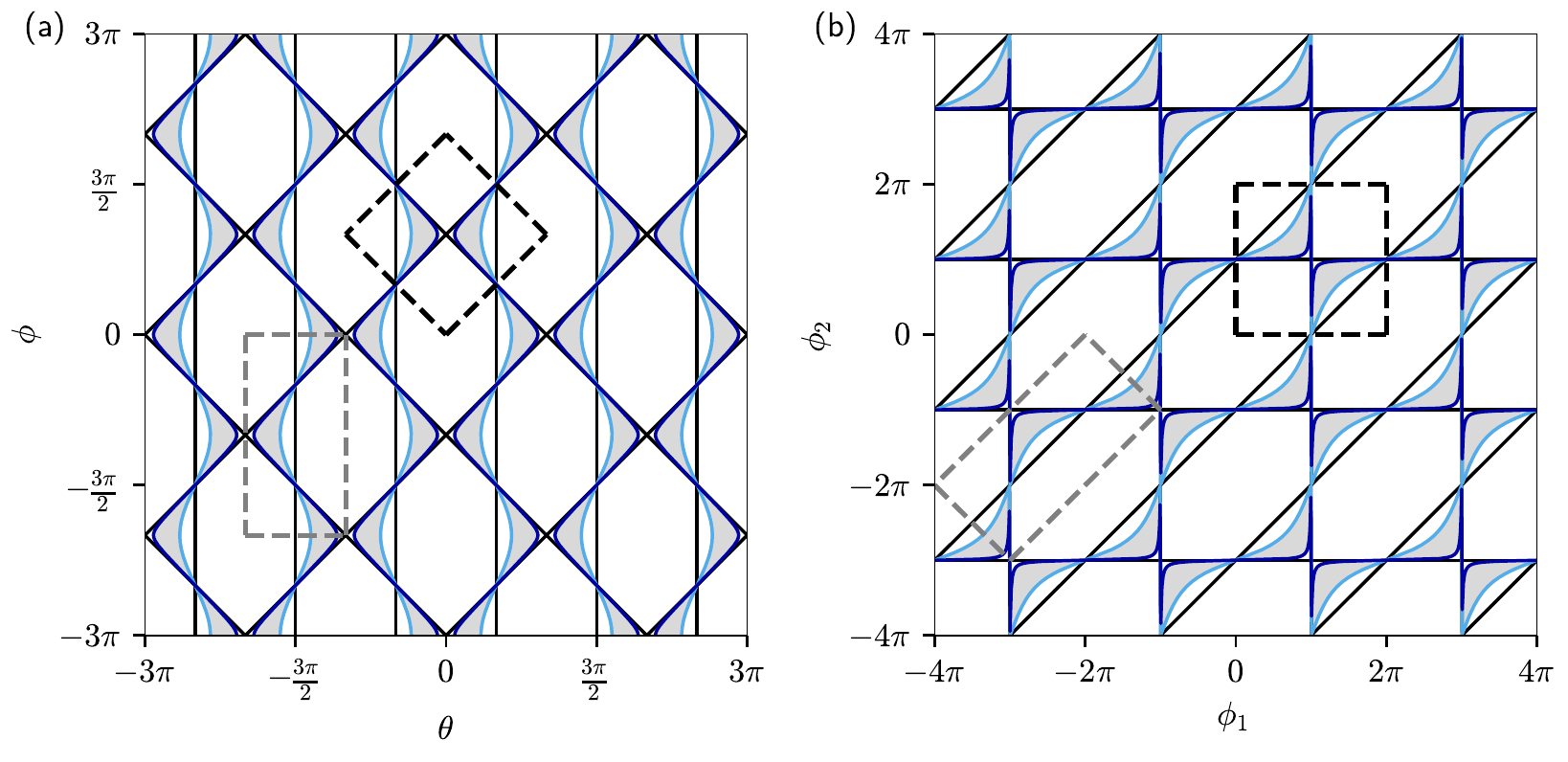}
    \caption{(a)~Phase diagram for the SNSNS geometry, as in Fig.~\ref{fig:SNSNS_junction} of the main text, with extended regions of the phases $\theta$, $\phi$.
    (b)~The phase diagram as a function of the parameters $\phi_1=\phi+\theta$, $\phi_2=\phi-\theta$. The periodic pattern appearing in the phase diagram is unique to the the phase-induced setup. The dashed black squares denote unit cells and the dashed gray rectangles the section of the phase diagram we plot in Fig.~\ref{fig:exact_PD_SNSNS} and Fig.~\ref{fig:TB} of the main text.
    \label{fig:extended_PD}}
\end{figure}

In Fig.~\ref{fig:extended_PD} we show the resulting phase diagram with extended regions of $\theta$ and $\phi$. This is the repeated version of Fig.~\ref{fig:exact_PD_SNSNS} of the main text, exposing the inherent periodicity of the phase diagram. Such a periodic feature, if observed in an experiment, could greatly support a topological interpretation of the data.

\section{Phase winding condition}
In Ref.~\cite{lesser_three-phase_2021}, an algebraic condition for the existence of a phase winding, i.e., a vortex, was derived: the quantity $f=\cos\left(2\theta\right)+\cos\left(\phi+\theta\right)+\cos\left(\phi-\theta\right)$ must be smaller than $-1$. Let us verify that this indeed holds in our system. We define $T=\tanh\left(W_{\rm S}/\xi\right)$ and calculate
\begin{equation}\label{seq:theta_phi_relation}
    \begin{aligned}f & =\cos\left(2\theta\right)+\cos\left(\phi+\theta\right)+\cos\left(\phi-\theta\right)\\
 & =-1+2\cos^{2}\left(\theta\right)+2\cos\left(\phi\right)\cos\left(\theta\right)\\
 & =-1+2T^{2}\cos^{2}\left(\phi\right)-2T\cos^{2}\left(\phi\right)\\
 & =-1-\underbrace{2T\left(1-T\right)\cos^{2}\left(\phi\right)}_{>0}\\
 & <-1,
\end{aligned}
\end{equation}
since $0<T<1$.

Moreover, the condition for a phase winding may be further simplified using some trigonometric identities:
\begin{equation}
    \begin{aligned}0 & < 1 + \cos\left(2\theta\right)+\cos\left(\phi+\theta\right)+\cos\left(\phi-\theta\right)\\
 & = 2\cos^2\left(\theta\right) + 2\cos\left(\theta\right)\cos\left(\phi\right) \\
 & = 2\cos\left(\theta\right) \left[ \cos\left(\theta\right) + \cos\left(\phi\right) \right].
\end{aligned}
\end{equation}
We have the freedom to choose $-\pi/2\leq\theta\leq\pi/2$, which implies $\cos\left(\theta\right)\geq0$. We may then eliminate this factor to find the simple condition referred to in the main text,
\begin{equation}
    \cos\left(\theta\right) + \cos\left(\phi\right) < 0.
\end{equation}

\section{Optimization of parameters}

\label{App:opt}

Inspecting Eq.~\eqref{seq:theta_phi_relation} (or equivalently Eq.~\eqref{eq:zero_E_condition} of the main text), and the corresponding phase diagram (Fig.~\ref{fig:exact_PD_SNSNS} of the main text), we find that the topological region is the area between two implicitly determined curves.
We now ask the question of how this area can be optimized, given certain practical degrees of freedom.

We begin by calculating the area of the topological phase in the $\theta$--$\phi$ plane.
Due to the symmetry of the curves, it is enough to consider just $\phi\in\left[\frac{\pi}{2},\pi\right]$.
The largest possible topological area in this region is determined by the triangle defining the existence of a vortex (or, equivalently, infinite velocity imbalance):
\begin{equation}
    S_{\rm max} = \frac{1}{2} \times \frac{\pi}{2} \times \frac{\pi}{2} = \frac{\pi^2}{8}.
\end{equation}
The area of the actual topological phase is given by
\begin{equation}\label{seq:area_integral}
    S = \int_{\frac{\pi}{2}}^{\pi} d\phi \left\{ \cos^{-1}\left[ \tanh\left(\frac{W_{\rm S}}{\xi_{<}}\right) \cos\left(\phi\right)\right] - \cos^{-1}\left[ \tanh\left(\frac{W_{\rm S}}{\xi_{>}}\right) \cos\left(\phi\right) \right] \right\}.
\end{equation}
Here $\xi_{<}$ is the smaller coherence length corresponding to the smaller velocity $v_{<}$, and $\xi_{>}$ is the larger coherence length corresponding to the larger velocity $v_{>}$.
While this integral's analytical form is not very revealing, it is readily computed numerically.

\begin{figure}
    \centering
   \includegraphics[width=\linewidth]{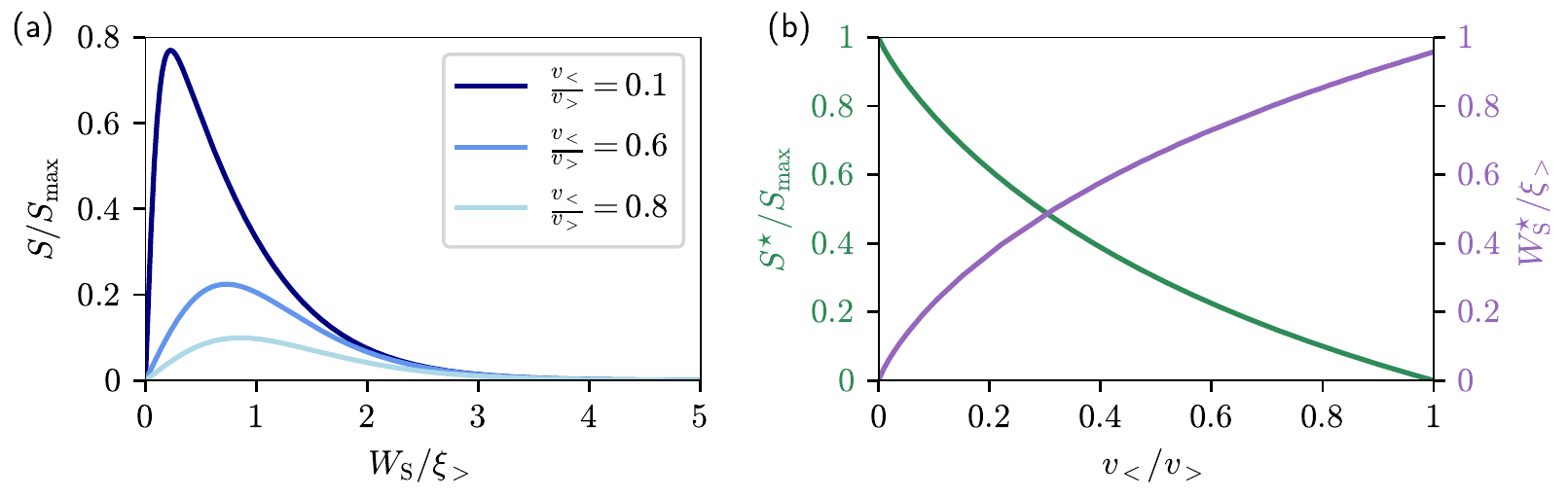}
    \caption{Optimization of the topological area in parameter space as a function of the velocity imbalance and middle superconductor width. 
    (a)~Area of the topological region in $\theta$--$\phi$ space $S$, normalized by its maximal possible value $S_{\rm max}$ (the triangles bordered by black lines in Fig.~\ref{fig:exact_PD_SNSNS} of the main text), as a function of the width of the middle superconductor $W_{\rm S}$, normalized by the larger coherence length $\xi_{>}$. Three curves are shown for different values of the ratio between the smaller velocity $v_{<}$ and the larger one $v_{>}$.  
    (b)~For each value of the ratio $v_{<}/v_{>}$ we find the optimal value of $W_{\rm S}$ (shown in purple), yielding the largest topological area (shown in dark green). This value is always intermediate between $\xi_{<}$ and $\xi_{>}$.
    \label{sfig:topo_area}}
\end{figure}

For a given velocity ratio $v_{<}/v_{>}$, we plot the dependence of the topological area $S$ on the width of the middle superconductor $W_{\rm S}$ in Fig.~\figref{sfig:topo_area}{a}. 
We observe a non-monotonic behavior: the topological area approaches zero at both limits $W_{\rm S}\to0$ and $W_{\rm S}\to\infty$, reaching an optimal value $S^{\star}$ at some intermediate value $W_{\rm S}^{\star}$. 
Both $S^{\star}$ and $W_{\rm S}^{\star}$ depend on the velocity ratio $v_{<}/v_{>}$.
In Fig.~\figref{sfig:topo_area}{b} we plot their dependence on $v_{<}/v_{>}$, finding monotonic and opposite behavior for the two quantities.
As expected, the optimal topological area $S^{\star}$ goes from zero when the two velocities are identical to $S_{\rm max}$ when they are very different.
The optimal width $W_{\rm S}^{\star}$ increases when the two velocities are close to one another, and is relatively small when they are very different.
Given the velocity imbalance associated with a particular material at a specific orientation, this calculation provides a tool for optimizing the device's geometry.

\section{Velocity imbalance in a nearest-neighbor Rashba square lattice}

In this section we show how the Fermi velocity imbalance appears in a Rashba square lattice with nearest-neighbor hopping only, in particular directions. This is a special case
of the general discussion of the main text, see Eq.~\eqref{eq:H2D}.

In a 1D model constrained to have only nearest-neighbor tunneling, even the most general spin-orbit coupling will eventually lead to the two spectra $E_\pm(k) = E (k \pm k_{\rm SO})$ with $k_{\rm SO}$ being a constant~\cite{meir_universal_1989}.
The velocities of the two branches defined by $v_\pm =\partial E_\pm( k) /\partial k$ are thus identical. To get different velocities it is necessary to get an effective 1D model with next-nearest-neighbor coupling. We show now that when choosing the angle $\theta$ between the superconducting pads and the square lattice's primitive vectors properly, a velocity imbalance occurs naturally. 

The Hamiltonian of a square lattice with Rashba spin-orbit coupling is given by the Hamiltonian in Eq.~\eqref{eq:H2D}, with $\lambda_z=0, t_i = t \cos (k_{\rm SO} a), \lambda_x=\lambda_y = t \sin( k_{\rm SO} a)$ yielding 

\begin{equation}
    H=t\sum_{i=1,2}\cos \left[({\bf k}-{k}_{\rm SO}{\hat z}\times {\bf \sigma})\cdot {\bf a}_i \right]
\end{equation}
where ${\bf a}_{1,2}$ are the basis vector spanning the square lattice.
Choosing the direction of the superconducting pads in an angle $\theta$ with respect to the direction of the primitive square lattice vector, we obtain the effective 1D model in the SNSNS ($\hat y$) direction to be
\begin{equation}
    H=t\cos \left({k}a\cos{\theta}-{k}_{\rm SO}a\sigma_y\right)
    +t\cos \left({k}a\sin{\theta}+{k}_{\rm SO}a\sigma_y\right)
    \label{Rashbaeff1D}
\end{equation}

Eq.~\eqref{Rashbaeff1D} is a 1D hopping Hamiltonian. For $\theta=0,\pi/2$ this Hamiltonian allows only for nearest-neighbor hopping, leading to a spectrum of two rigidly shifted cosine bands, with no imbalance in the velocities. 

For any rational value of $\tan\theta=m/n$, Eq.~\eqref{Rashbaeff1D} defines a unit cell of size ${\tilde a}=a/\sqrt{m^2+n^2}$, leading to $ka\cos{\theta}=k{\tilde a}m$ and $ka\sin{\theta}=k{\tilde a}n$. With these observations, Eq.~\eqref{Rashbaeff1D} becomes a 1D Hamiltonian with an $m$-site hopping term where the SOC term is proportional to $\sigma_x$, and an $n$-site hopping term where the SOC term is proportional to $\sigma_y$. Any pair of values $n\ne m$ allows for hopping loops in which SOC leads to the accumulation of an Aharonov--Casher phase, which modifies the spectrum beyond a rigid shift. 

\begin{figure}
    \centering
   \includegraphics[width=0.7\linewidth]{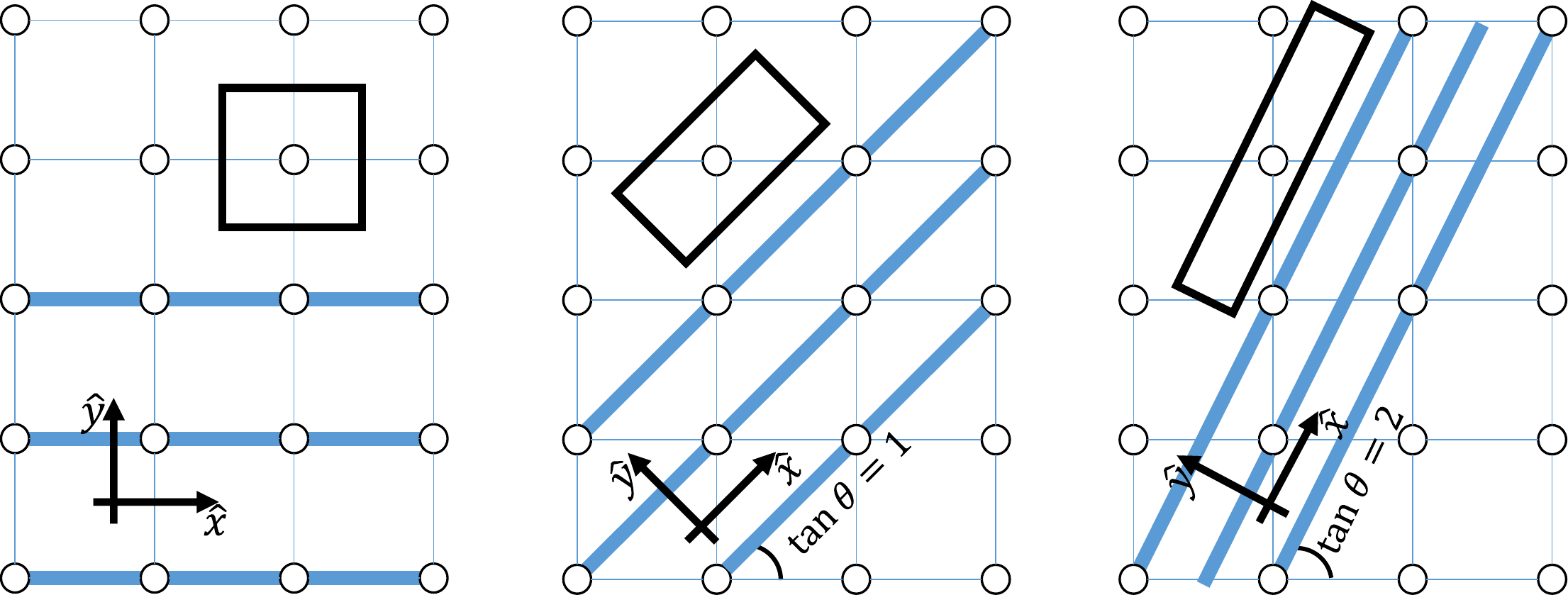}
    \caption{The superconducting pads are aligned at different orientations $\theta$ with respect to the underlying square lattice. In the effective 1D model, next-nearest-neighbor hopping (in the $y$ direction, perpendicular to the bold lines) is generated unless $\theta$ is an integer multiple of $\pi/4$. The black squares denote the unit cells. 
    \label{sfig:nn}}
\end{figure}

Fig.~\ref{sfig:nn} shows two cases in which the 1D model has nearest-neighbor coupling only, namely $m=0$ and $m=n=1$, as well as a third case, where $m/n=1$, in which two-site hopping is enabled, with a resulting imbalance in the Fermi velocities. 

\section{Velocity imbalance from next-nearest-neighbor hopping in 1D}

In this section we show how a Fermi velocity imbalance appears as a consequence of next-nearest neighbor hopping in a Rashba band, even in directions at which the Fermi velocities of the two branches are equal for nearest-neighbor hopping. 

We begin from the one-dimensional tight-binding model introduced in the main text, with $k_x=0$:
\begin{equation}\label{seq:nnn_1D}
    H(k_y) = 2t\left[1-\cos\left(k_y\right)\right] + 2t^{\prime}\left[1-\cos\left(2k_y\right)\right] + 2\alpha\sin\left(k_y\right)\sigma_x - \mu,
\end{equation}
where $t,t^{\prime}$ are the nearest and next-nearest hopping amplitudes, respectively, $\alpha$ is the Rashba SOC amplitude, and $\mu$ is the chemical potential.
Since $\sigma_x$ is a good quantum number, we can write the spectrum as 
\begin{equation}
    E_{\sigma_x}(k_y) = -2\tilde{t}\cos\left( k_y + \lambda\sigma_x \right) + 2t^{\prime}\left[1-\cos\left(2k_y\right)\right] - \tilde{\mu},
\end{equation}
where $\tilde{t}=\sqrt{t^2+\alpha^2}$, $\sin\lambda=\alpha/\tilde{t}$, and $\tilde{\mu}=\mu-2t$. 
In this form it is evident that without $t^{\prime}$, the spectrum is composed of two  cosines that are shifted in momentum with respect to one another, resulting in identical Fermi velocities. This is the case of $m=0$ of the previous subsection. 

To demonstrate the importance of $t^{\prime}$, let us consider the simple case $t=\alpha$ such that $\lambda=\pi/4$, and $\tilde{\mu}=2t^{\prime}$. The spectrum is then 
\begin{equation}\label{eq:nnn_sweet_spot}
    E_{\pm}(k_y) = -2\tilde{t}\cos\left( k_y \pm \frac{\pi}{4} \right) + 2t^{\prime}\cos\left(2k_y\right),
\end{equation}
and then the Fermi points for $E=0$ are $k_{y+}=\pi/4$, $k_{y-}=3\pi/4$. Differentiating Eq.~\eqref{eq:nnn_sweet_spot} with respect to $k_y$ and substituting these values, we find that the velocities are 
\begin{equation}
    v_{\pm} = 2\left( \tilde{t} \pm 2t^{\prime} \right).
\end{equation}
We conclude that $t^{\prime}\neq0$ is necessary to induce a velocity imbalance in this direction.

\begin{figure}
    \centering
   \includegraphics[width=\linewidth]{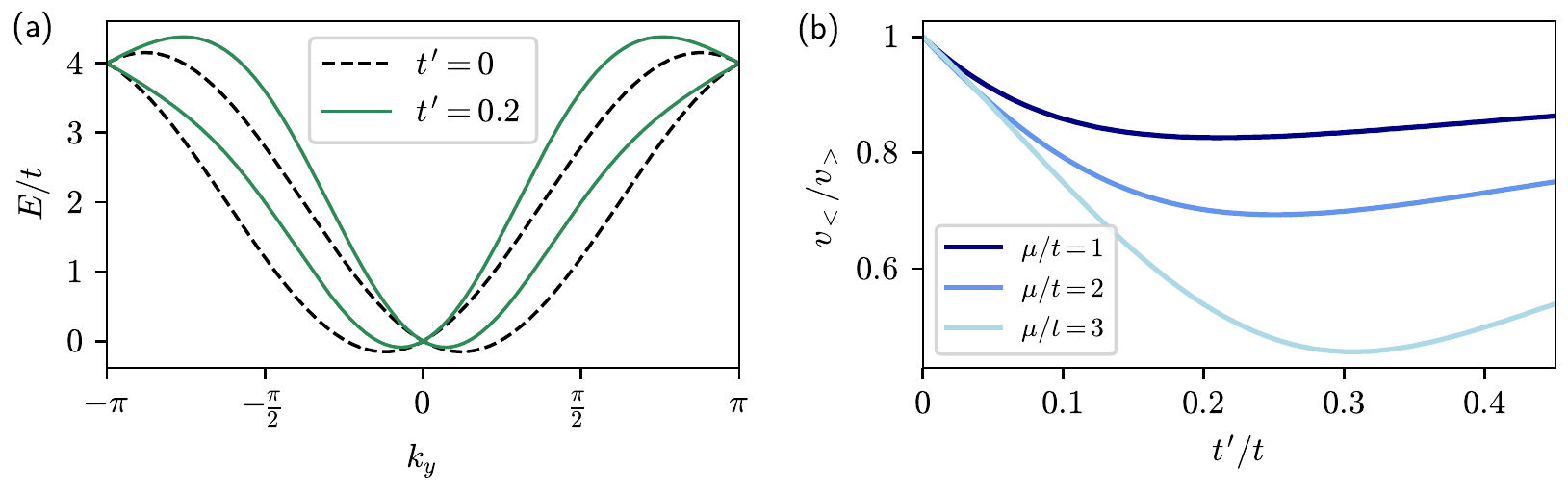}
    \caption{Effect of next-nearest neighbor hopping on the Fermi velocity imbalance. 
    (a)~Spectrum of a one-dimensional lattice model with nearest neighbor hopping amplitude $t=1$, Rashba spin-orbit coupling amplitude $\alpha=0.4$, and next-nearest neighbor hopping amplitude $t^{\prime}$, see Eq.~\eqref{seq:nnn_1D} of the main text. Dashed black lines correspond to $t^{\prime}=0$, where the spectrum is simply two shifted cosines with equal Fermi velocities regardless of the Fermi energy. Solid green lines correspond to $t^{\prime}=0.2$, where a Fermi velocity imbalance is induced.
    (b)~Velocity imbalance (ratio between the smaller velocity $v_{<}$ and the larger one $v_{>}$) as a function of the next-nearest hopping amplitude $t^{\prime}$, for several values of the chemical potential $t$. For a stable topological phase, significant deviation from $v_{<}/v_{>}=1$ is favorable. This is obtained at an intermediate value of $t^{\prime}$.
    \label{sfig:nnn_effect}}
\end{figure}

To show this more generally, we numerically solve the problem. The spectrum of the Hamiltonian (at $\mu=0$) is shown in Fig.~\figref{sfig:nnn_effect}{a}: for $t^{\prime}=0$ we obtain two cosines shifted from zero by the spin-orbit coupling, whereas for $t^{\prime}\neq0$ they become distorted and are no longer pure cosines.

The distortion induced by the next-nearest hopping is examined in Fig.~\figref{sfig:nnn_effect}{b}. 
For several values of the chemical potential, we numerically find the Fermi velocities.
To do this, we find the four Fermi points by equating the eigenvalues of the Hamiltonian~\eqref{seq:nnn_1D} to zero.
Then, we numerically differentiate the dispersion at the Fermi points.
Focusing on $k_y>0$, we find two velocities, and calculate the velocity imbalance as the ratio of the smaller one to the larger one.

The velocity imbalance depends non-monotonously on $t^{\prime}$, as seen in Fig.~\figref{sfig:nnn_effect}{b}.
Recall that for obtaining a stable topological phase, a small value of $v_{<}/v_{>}$ is needed [see Fig.~\figref{sfig:topo_area}{b}]. 
We find that this happens at an intermediate value of $t^{\prime}$.
Furthermore, in this model, it is preferred to increase the chemical potential, since that decreases the optimal $v_{<}/v_{>}$.

\begin{figure}[b]
    \centering
   \includegraphics[width=\linewidth]{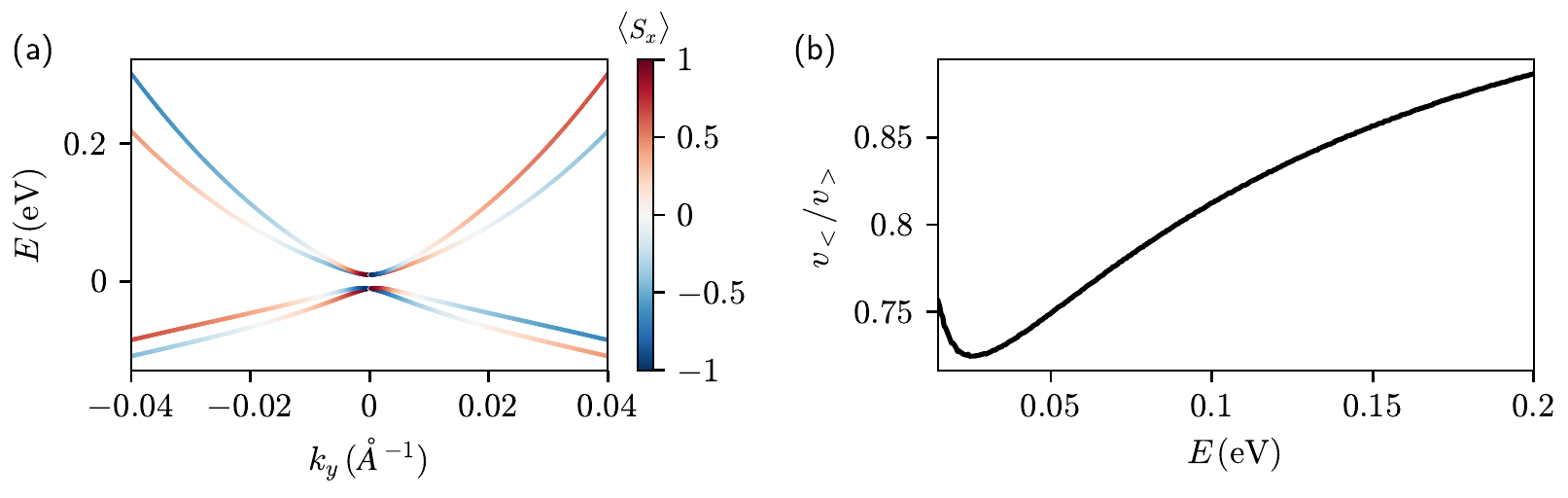}
    \caption{Fermi velocity imbalance in HgTe. 
    (a)~Spectrum of a gate-defined wire in a HgTe 2D electron gas, simulated according to Ref.~\cite{reuther_gate-defined_2013}. Colors indicate the spin projection along the wire's direction $x$, and $k_y$ is the momentum transverse to the wire.
    (b)~Velocity imbalance $v_{<}/v_{>}$ as a function of the Fermi energy. We observe a non-monotonic behavior with regions of substantial velocity imbalance.
    \label{sfig:HgTe}}
\end{figure}

To be more concrete, we demonstrate this effect in an actual material, HgTe, for which a gate-defined wires may be engineered~\cite{reuther_gate-defined_2013}.
The spectrum in the direction transverse to the wire is shown in Fig.~\figref{sfig:HgTe}{a}.
The combination of the Dirac-like band structure, terms quadratic in momentum, and Rashba spin-orbit coupling leads to different Fermi velocities for the two spin branches.
Figure~\figref{sfig:HgTe}{b} shows the velocity imbalance as a function of the Fermi energy for this configuration.
We observe Fermi energies for which the velocity imbalance is significant, and these can potentially host a topological phase.

\section{Velocity imbalance from a periodic potential along the junction}
\label{App:periodic_potential}
In this section we demonstrate another mechanism for inducing an imbalance in the transverse Fermi velocities, leading to a topological phase transition, using a periodic potential.
We consider three superconducting slabs, each of width $W$, whose superconducting phases are $-\theta$, $\phi$, $\theta$:
\begin{equation}
    \Delta\left(y\right)=\begin{cases}
\Delta e^{-i\theta}, & -\frac{3W}{2}\leq y<-\frac{W}{2}\\
\Delta e^{i\phi}, & -\frac{W}{2}\leq y<\frac{W}{2}\\
\Delta e^{i\theta}, & \frac{W}{2}\leq y\leq\frac{3W}{2}.
\end{cases}
\end{equation}
The superconductors are proximity coupled to a 2D electron gas with Rashba spin-orbit coupling. The normal Fermi surface then has equal Fermi velocities for the two branches. 

In order to make the Fermi velocities distinguish between the branches, we add a periodic potential in the $x$ direction. The rationale of doing so is that while at $k_x=0$ spin-orbit coupling may be gauged away, and hence does not affect the spectrum of sub-gap states, a periodic potential would mix into the $k_x=0$ eigenstates components of $k_x=\pm nq$, in whish spin-orbit coupling affects the spectrum. 

\begin{figure}
    \centering
   \includegraphics[width=\linewidth]{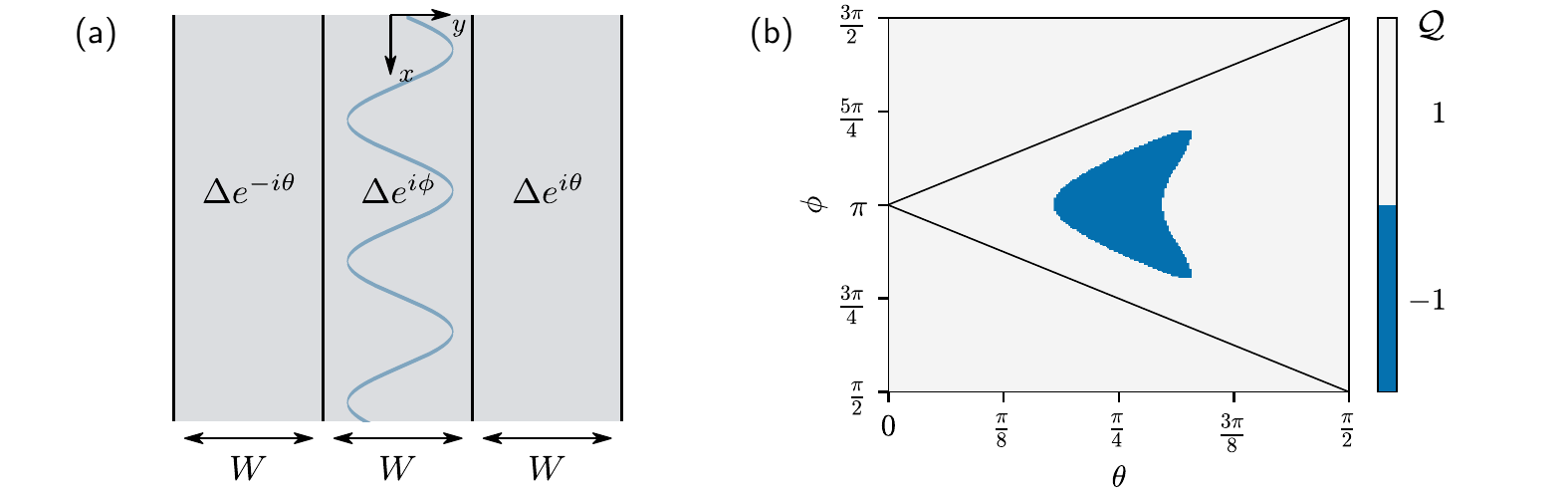}
    \caption{Periodic potential along the junction as a mechanism enabling topological superconductivity.
    (a)~Model system: three superconducting pads with distinct phases, on top of a spin-orbit-coupled 2D electron gas, with a periodic modulation of the chemical potential in the middle superconductor (illustrated by the oscillating blue curve); see Eq.~\eqref{seq:H_periodic_potential}.
    (b)~Topological phase diagram of the model in the $\theta$--$\phi$ plane. Colors indicate the topological invariant $\cal {Q}$, which is $+1$ (gray) in the trivial phase and $-1$ (blue) in the topological phase. Notice the similarity to the analytically obtained phase diagram Fig.~\ref{fig:exact_PD_SNSNS} of the main text.
    The parameters used are $W=4$, $t=1$, $\Delta=0.7$, $\alpha=0.35$, $\mu_0=1$, $\delta\mu=-0.8$, $q=1/5$.
    \label{sfig:periodic_potential}}
\end{figure}

Thus, we introduce a periodic potential in terms of a modulation of the chemical potential in the middle superconductor, such that 
\begin{equation}
    \mu(x,y) = \begin{cases}
\mu_{0}, & -\frac{3W}{2}\leq y<-\frac{W}{2}\ \text{or }\frac{W}{2}\leq y\leq\frac{3W}{2}\\
\mu_{0}+\delta\mu\cos\left(2\pi qx\right), & -\frac{W}{2}\leq y<\frac{W}{2}.
\end{cases}
\end{equation}
The model is illustrated in Fig.~\figref{sfig:periodic_potential}{a}.
The full real-space Hamiltonian is given by 
\begin{equation}\label{seq:H_periodic_potential}
    \begin{aligned}H & =\sum_{x,y}\left[\mu\left(x,y\right)c_{x,y}^{\dagger}\sigma_{0}c_{x,y}+\Delta\left(y\right)c_{x,y,\uparrow}^{\dagger}c_{x,y,\downarrow}^{\dagger}\right.\\
 & \left.+c_{x,y}^{\dagger}\left(t\sigma_{0}+i\alpha\sigma_{y}\right)c_{x+1,y}+c_{x,y}^{\dagger}\left(t\sigma_{0}-i\alpha\sigma_{x}\right)c_{x,y+1}+\text{H.c.}\right].
\end{aligned}
\end{equation}

To study the topological properties of this model, we Fourier-transform the Hamiltonian in the $x$ direction.
For rational $q=q_{\rm N}/q_{\rm D}$, the unit cell contains $q_{\rm D}$ sites along $x$ (multiplied by $3W$ sites along $y$). 
We may then calculate the momentum-space topological invariant~\cite{kitaev_unpaired_2001,altland_nonstandard_1997,schnyder_classification_2008,wimmer_algorithm_2012}, 
\begin{equation}\label{eq:pfaffian}
    \mathcal{Q} = {\rm sign} \left[ {\rm Pf}\left( \mathcal{P} \mathcal{H}(k_{x}=0) \right) {\rm Pf}\left( \mathcal{P} \mathcal{H}(k_{x}=\pi) \right) \right],
\end{equation}
where ${\rm Pf}(\cdot)$ is the Pfaffian and $\mathcal{P}$ is the particle-hole operator.
${\cal Q}$ is equal to $+1$ in the trivial phase and $-1$ in the topological phase.
The numerically calculated phase diagram is shown in Fig.~\figref{sfig:periodic_potential}{b}. 
The phase diagram indeed supports topological regions, and as expected they only appear when phase winding is present. Notice that Fig.~\figref{sfig:periodic_potential}{b} is a proof of principle, where no comprehensive attempt has been made to optimize the choice of the periodic potential. 

\section{Velocity imbalance in a two-band system}

\label{App:Twobands}

\begin{figure}
    \centering
   \includegraphics[width=0.5\linewidth]{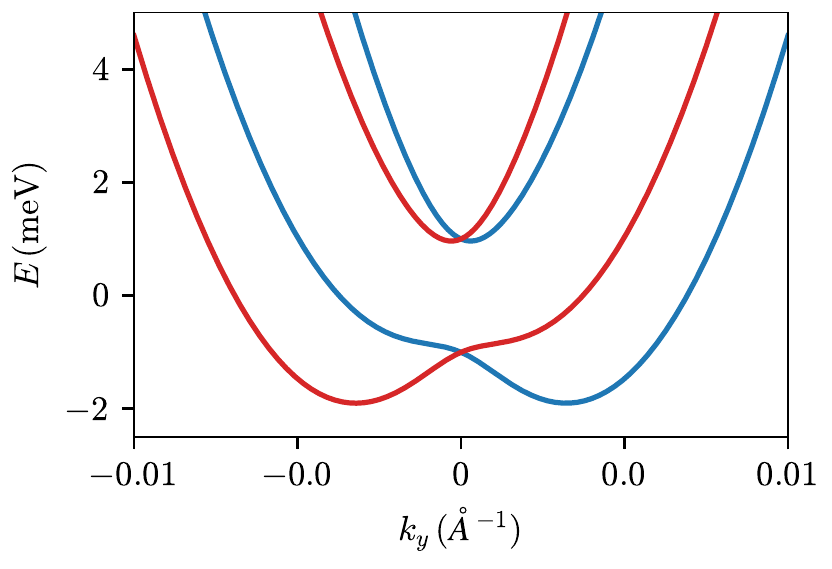}
    \caption{\label{fig:alphaz}The two-band model of Eq.~\eqref{eq:supp_two_bands} [see also Eq.~\eqref{eq:alphaz} of the main text], with appropriate parameters for InAs ($m^{*}=0.026m_e$). The bare spin-orbit coupling is $\alpha=10\,{\rm meV nm}$, and we take the the spin-orbit coupling at the bottom layer to be $-\frac{1}{2}$ that of the top layer. This results in $\alpha_{11}=\alpha_{22}=2.5\,{\rm meV nm}$ and $\alpha_{12}=7.5\,{\rm meV nm}$. We take the inter-layer coupling $t=1\,{\rm meV}$ , so that $\delta E=2t=2\,{\rm meV}$. The velocity imbalance is most pronounced between $-1.5\,{\rm meV}$ and $-0.5\,{\rm meV}$. Colors indicate spin up (red) and down (blue).}
\end{figure}

In Fig.~\ref{fig:alphaz} we plot the spectra of the eigenvalues of Eq.~\eqref{eq:alphaz} of the main text, for typical parameters of InAs, to establish the correspondence to Ref.~\cite{lesser_three-phase_2021}. There, two layers of a spin-orbit-coupled semiconductor are coupled with tunneling amplitude $t$. We can start without the spin-orbit coupling and diagonalize the Hamiltonian to find the bands --- these are the symmetric and anti-symmetric superpositions of wavefunctions of the two layers. There is an energy splitting $\delta E=2t$ between these two bands. We then add the spin-orbit coupling, and in the basis of the bands, the transverse Hamiltonian (at $k_x=0$) takes the form
\begin{equation}\label{eq:supp_two_bands}
H\left(k_y\right) = \begin{pmatrix}\frac{k_y^{2}}{2m^{*}} & \alpha_{11}k_y & 0 & \alpha_{12} k_y\\
\alpha_{11}k_y & \frac{k_y^{2}}{2m^{*}} & \alpha_{12} k_y & 0\\
0 & \alpha_{12} k_y & \frac{k_y^{2}}{2m^{*}}+\delta E & \alpha_{22}k_y\\
\alpha_{12} k_y & 0 & \alpha_{22}k_y & \frac{k_y^{2}}{2m^{*}}+\delta E
\end{pmatrix},
\end{equation}
where $m^{*}$ is the effective mass. Diagonalizing this Hamiltonian, we find indeed two different Fermi velocities, depending on the position of the chemical potential. This is shown for typical quantum well parameters in Fig.~\ref{fig:alphaz}.
Notice that the tunneling amplitude $t$ should be of the same order as the spin-orbit energy to get an appreciable effect. This implies that one should use either wide quantum wells or double wells.

\end{document}